\documentclass[journal]{IEEEtran}
\usepackage{cite}

\ifCLASSINFOpdf
  \usepackage[pdftex]{graphicx}
  \graphicspath{ {./images/} }
  \DeclareGraphicsExtensions{.pdf,.jpeg,.png}
\else

\fi
\usepackage{array}
\usepackage{multirow}
\usepackage{tabularx}
\usepackage{multirow}
\usepackage{graphicx}
\usepackage[normalem]{ulem}
\usepackage{MnSymbol}

\usepackage{booktabs,caption}
\usepackage[flushleft]{threeparttable}

\useunder{\uline}{\ul}{}

\begin{document}

\title{Post-Quantum Cryptography for Internet of Things: A Survey on Performance and Optimization}
%
%

\author{Tao~Liu,
          Gowri~Ramachandran,
          and~Raja~Jurdak}%

%
%

\maketitle

\begin{abstract}

Due to recent development in quantum computing, the invention of a large quantum computer is no longer a distant future. Quantum computing severely threatens modern cryptography, as the hard mathematical problems beneath classic public-key cryptosystems can be solved easily by a sufficiently large quantum computer. As such, researchers have proposed PQC based on problems that even quantum computers cannot efficiently solve. Generally, post-quantum encryption and signatures can be hard to compute. This could potentially be a problem for IoT, which usually consist lightweight devices with limited computational power. In this paper, we survey existing literature on the performance for PQC in resource-constrained devices to understand the severeness of this problem. We also review recent proposals to optimize PQC algorithms for resource-constrained devices. Overall, we find that whilst PQC may be feasible for reasonably lightweight IoT, proposals for their optimization seem to lack standardization. As such, we suggest future research to seek coordination, in order to ensure an efficient and safe migration toward IoT for the post-quantum era.

\end{abstract}


\IEEEpeerreviewmaketitle

\section{Introduction}

\IEEEPARstart{I}{nternet} of Things (IoT) is a promising technology that emerged in recent years. IoT has many potential applications, ranging from connected personal gadgets to large-scale industrial drone networks. It has been estimated that 12.2 billion IoT devices were active in 2021, and this number will grow to 27 billion by 2025\cite{Hasan2022}. The security of IoT technologies remains a concern, as the computation cost of cryptography algorithms used for digital encryption and authentication may be burdensome for resource-constrained IoT node devices that were designed to be deployed en masse, and sometimes in inaccessible remote areas. Very often, factors such as production cost, size and battery life are prioritized over security for such low-end lightweight devices. If IoT technologies are indeed to integrate with agriculture, industrial production, medical service, military and everyday lives in general, adequate digital encryption and authentication standards should be enforced. Currently, one significant issue with applying cryptography in IoT is their high computational cost, which often scales with the key sizes of algorithms used. 

On the other hand, an upgrade for cryptographic standards will soon be necessary due to the latest development in quantum computing, as this advanced computation technology may force the world to use even more complex cryptographic algorithms, further exacerbating the current problem for IoT systems. A sufficiently large quantum computer can use Shor’s algorithm \cite{Shor1994} to solve Integer Factorization and Discrete Logarithmic problems, both used as fundamental hard problems for today's public-key cryptosystems. Additionally, Grover’s algorithm \cite{Grover1996} makes symmetric keys easier to search by providing quadratic speed up for the unstructured search problem. As such, whilst symmetric ciphers need to double the key size in order to maintain their security level, asymmetric cryptography must undergo a fundamental change in response to the emergence of quantum computing. These developments threaten the security of contemporary cryptographic schemes used in IoT applications.Formally, cryptography that is secure against an attacker with large quantum computers is categorized as Post-Quantum Cryptography (PQC).

Naturally, to ensure security in  future IoT systems, lightweight devices should also be protected by PQC. Yet a very practical problem remains: in general, quantum-resistant encryption and signature schemes are more computationally intensive than any public-key cryptosystems currently in-use \cite{Malina2021}. As a result, the migration toward PQC may have an undesired impact on IoT systems, where devices typically have limited energy supply, memory sizes, and processing speed. Recently, such complications have been noticed by researchers, and efforts were made to evaluate and optimize the performance of PQC algorithms in resource constrained devices. A comprehensive survey on this topic is \cite{8932459}, underscoring the problem that complex PQC algorithms may not be efficient enough on resource constrained IoT devices. This work reviews typical IoT network structures and past efforts for PQC migration. Furthermore, it surveys literature on the performance of candidates of a major PQC standardization project \cite{nist-pqc-proj} led by the United States National Institute of Standards and Technology (NIST), and compares their suitability for IoT systems. The content of this survey remains largely relevant today, but more recent developments warrant revisiting its findings. For example, this survey concluded that few works have proposed Post-Quantum solutions specific to resource-constrained IoT devices, but recent projects, such as pqm4 \cite{PQM4}, have gained traction in this direction since the survey's publication. We also noticed that some algorithms recommended by this survey were eliminated in the NIST PQC standardization project, rendering them less relevant in today's context. 

There are also other works that investigate the Post-Quantum security issues for IoT. Malina et al. \cite{9363165} is a survey on privacy-protecting technologies for IoT systems. Although it surveys the performance of several round-3 NIST PQC candidates, we found the information it provides is not sufficiently up-to-date: for example, NIST later selected Kyber and Dilithium as the primary KEM and signature schemes respectively \cite{nist-third-round}, but the article only includes performance evaluation for one variation from each algorithm: Kyber1024 and Dilithium-III. Furthermore, data for both algorithms were drawn from sources published before 2019. The authors of \cite{seyhan2021}, on the other hand, review the security requirement at different layers of IoT and the application of lattice-based cryptography (LBC) for IoT security. The authors further propose a classification framework for the application of LBC in resource-constrained devices. However, an IETF classification for resource-constrained devices published in 2014 \cite{ieft-rc-classification} was used, and the paper concludes that LBC is inefficient in many use cases. Such assessment may have under-evaluated, as many legacy devices originally covered by the IETF classification (such as 8-bit microprocessors) have become less relevant today. 

In light of the recent NIST decisions on the first several PQC algorithms to be standardized, PQC for IoT needs to be revisited. A survey which reviews the most recent development on PQC for IoT security is needed. Additionally, as we explained earlier, we believe an emphasis should be placed on the performance evaluation and optimization of PQC algorithms for resource constrained devices. The specific contributions of this paper are as follows:

\begin{enumerate}
    \item we evaluate recent works on PQC optimisation and evaluation.
    \item we review the current trends and identify the emerging gaps in this research area.
    \item we discuss potential future directions for the application of (optimized) PQC for IoT. 
\end{enumerate}  

The rest of this paper is structured as follows: in section II, we discuss preliminary knowledge required for further discussion of the topic, including the application of cryptography, quantum threat against modern cryptography, taxonomy of post-quantum cryptography and the basic structure of IoT systems. Section III presents our methodology for selecting and reviewing literature. In section IV-VII, we divide key literature based on their topics and discuss recent research trends. Section VIII summarizes our findings and provides recommendations on future research direction. 

\section{Preliminary Knowledge}

\subsection {Data Encryption for IoT}

In order to investigate the feasibility of applying PQC in IoT, it is important to first identify the application of contemporary pre-quantum cryptography in IoT. Figure 1 demonstrates the basic IoT network structure: the design goal is to connect the “Things”, i.e., IoT sensor/actuator devices to the Internet but full-blown Internet protocols such as TCP/IP do not fit into the resource-constrained hardware they operate on. Therefore, a gateway is required as a medium between the “Things” and the Internet \cite{zolotova2015}. The node-to-gateway connection is often wireless, allowing adversaries in proximity to eavesdrop if the broadcast is not encrypted. On the other hand, connections beyond the gateway can support more functionality due to the availability of network stacks and increased device capacity. A gateway may connect to an edge device, a cloud server, or join a many-to-many network over protocols such as Message Queuing Telemetry Transport (MQTT). It is also easier to implement security measures at this level, although the resource constraints may still be an issue for gateway devices that are mobile or in remote areas where energy and/or bandwidth are scarce. In general, we assume that the computation cost of securely transferring data to and from gateways and nodes may be heavy in the post-quantum era.

To date, Zigbee and LoRaWAN are two common technologies used for device-to-gateway communication, whilst MQTT is a publish-subscribe protocol used for data transmission from the gateways. In Table \ref{table_cryptosystems}  we present their currently supported encryption methods, and possible upgrades for them to be Post-Quantum secure. It is noted that both Zigbee and LoRaWAN rely only on 128-bit AES encryption, and the use of pre-installed keys is always necessary, whilst MQTT can use Transport Layer Security (TLS) and any custom payload encryption at application level. One issue about key pre-installation is the balance between scalability and security: on one hand, key diversification is usually preferred to limit the impact of side channel attacks \cite{urien-lwig-security-classes-09} (i.e., compromising one key will only compromise the device that uses it). On the other hand, keeping track of pre-installed keys can also be burdensome considering an IoT application may use thousands of small devices. Malik et al. \cite{Malik2019} surveyed IoT key bootstrapping protocols based on public key cryptography (PKC) and summarized proposals into five categories: Raw Public Key, Certificate-Based, Identity-Based, Self-Certified and Certificate-less. However, for most of these sophisticated protocols to function, public-key cryptographic functions need to be executed frequently in both server and client devices. Therefore, especially under a post-quantum context, it would be crucial to understand if public-key cryptographic functions are feasible to run on resource-constrained devices. 

\begin{figure}
\centering
\includegraphics[width=3.4 in]{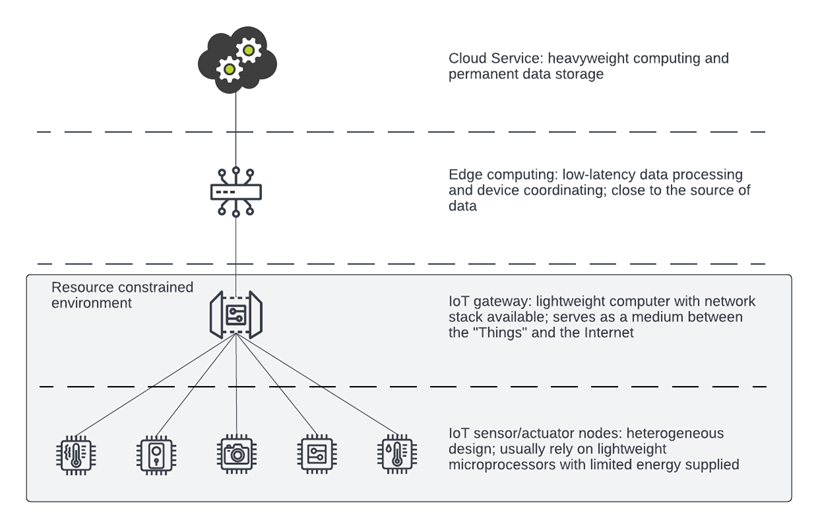}
\caption{Layers in a simple IoT network. In general, we consider IoT sensor/actuator node devices and gateway devices resource constrained. As such, two types of communication are resource constrained: the Node-to-Gateway layer and the Gateway-to-Internet layer.}
\label{fig_1}
\end{figure}

\subsection{Cryptography and Information Security}

Cryptography is closely associated with encryption. By using a key and an algorithm (consists of permutations, substitutions and other operations) to ``scramble'' a plaintext, it is possible to produce a ciphertext which can only be recovered with the correct key. Whilst the primary objective of cryptography is to protect \emph{confidentiality}, modern cryptography can be used to achieve three other objectives: data integrity, authentication and non-repudiation \cite{delfs2015introduction}. The usage of keys differentiates modern cryptography into two distinctive categories: symmetric cryptography and asymmetric cryptography (also known as public-key cryptography). In symmetric cryptography, the same key is used to both encrypt and decrypt data. In asymmetric cryptography, a ``private key'' is kept secret, and the ``public key'', as its name suggests, is made public. such arrangement allows interesting functionalities which is critical for information security: when the public key is used to encrypt, it ensures that a secret message can only be read by the intended recipient; when the secret key is used to encrypt, on the other hand, entities with the public key can verify that a non-repudiable message is indeed made by the secret key owner. 

Together, these two categories of cryptography guarantee the  secure transmission and storage of information in the digital world. One of the most important applications of cryptography is the Transport Layer Security (TLS) protocol, which is the standard method for establishing confidential and authenticated communication channels over the open Internet. With TLS, two distant parties can authenticate each other using digital signatures and third-party certificates, and then establish mutual secrets using asymmetric key exchange such as the Diffie-Hellman algorithm (without actually transmitting the secret, thus eliminating the risk of eavesdropping). Then, the mutual secret can be used as the key for symmetric ciphers, which are much more efficient for bulk data encryption and decryption.  

\subsection{Quantum Threats Against Cryptography}

It is known that quantum computing poses threats to both symmetric and asymmetric cryptography, but the threat against asymmetric cryptography is more significant \cite{Mavroeidis2018}. An overview of existing cryptosystems and how they are threatened by quantum computing is presented in Table \ref{table_cryptosystems}. As shown in the table, state-of-the-art asymmetric cryptography – including the RSA, the Elliptic-Curve Diffie-Hellman (ECDH) key exchange, and the Elliptic-Curve Digital Signature Algorithm (ECDSA) – offer no security against quantum computing. The security of RSA relies on the hardness of factoring large bi-prime numbers, also known as the Integer Factorization (IF) problem. On the other hand, Elliptic Curve Cryptography (ECC) takes the assumption that Elliptic Curve Discrete Logarithm (ECDL) problems are hard to solve. Integer Factorization and the Discrete Logarithm Problem are believed to be hard for classic computers, yet they can be solved in polynomial time by a quantum computer large enough to run Shor’s algorithm. Whilst small, experimental quantum computers today are incapable of solving practical ciphers, researchers have been estimating the quantum resources needed to achieve such a goal \cite{Baraban2010, Roetteler2017}. Other studies, such as \cite{Saxena2021, Bocharov2016} explore the possibility of using variations of Shor’s algorithm to factorize small bi-prime and tri-primes in quantum architectures. The quantum approach to solving the Discrete Logarithm problem over elliptic curves has also been discussed in theory in \cite{Larasati2021}. 

On attacking symmetric cryptography, Grover’s algorithm can potentially decrease the security strength of existing ciphers by offering a quadratic speed-up for exhaustive key-searching \cite{Grover1996}, but the practical implication of this attack is still debated as Grover’s algorithm requires queries to be run sequentially. Nevertheless, researchers have estimated that a quantum architecture with 2953, 4449, 6681 qubits can carry out Grover’s search against the Advanced Encryption Standard (AES) with 128-, 192- and 256-bit security, respectively \cite{Grassl2016}. Other works, such as \cite{10.1007/978-3-662-53008-5_8} and \cite{10.1007/978-3-030-34578-5_20}, outlines additional quantum attacks that can potentially reduce the hardness of exhaustive key-searching. 

Whilst theoretical estimations suggest at least several thousands of qubits are needed in a quantum computer to break existing cryptography, global technology giants are certainly making their efforts to reach this goal. Currently, IBM takes lead in quantum computing hardware with its 127-qubit processor \cite{IBM2021} and ambition to go beyond 1000 qubits in 2023\cite{IBM2020}. It has been predicted that the likelihood of having a scalable quantum computer in the next decade is significant \cite{Mosca2018}. As such, quantum-resistant encryption and signature schemes are urgently needed. 

\begin{table}[]
\centering
\caption{List of modern cryptosystems, their fundamental problems and security against quantum computing.}
\label{table_cryptosystems}
\begin{threeparttable}
    \begin{tabular}{>{\raggedright\arraybackslash}p{0.08\textwidth}|
                        >{\raggedright\arraybackslash}p{0.15\textwidth}|
                        >{\raggedright\arraybackslash}p{0.08\textwidth}|
                        >{\raggedright\arraybackslash}p{0.08\textwidth}}
    \hline
      Cryptosystem &  Fundamental Problem &  Current State &  Secure Against QC? \\ \hline\hline
      Diffie-Hellman (DH) &
      \multirow{2}{0.15\textwidth}{Discrete Logarithm} &
      Outdated &
      \multirow{5}{0.15\textwidth}{No} \\ \cline{1-1} \cline{3-3}
      DSA & & Outdated & \\ 
       \cline{1-3}
    RSA & Integer Factorization & In Use & \\ 
       \cline{1-3}
    ECDH &
      \multirow{2}{0.15\textwidth}{Discrete Logarithm over Elliptic Curves} &
      In Use & \\ 
       \cline{1-1} \cline{3-3}
    ECDSA &
       & In Use & \\ \hline
    Symmetric Ciphers &
      Exhaustive search \& collision-finding in large key spaces & In Use & Yes$^1$ \\ \hline 
      PQC &
      Lattice Problems, Error-Correction Codes, Multivariate Systems, Hash Trees, Isogeny Graphs &
      Developing &
      Yes \\ \hline
    \end{tabular}
    
    \begin{tablenotes}
        \small
        \item{$^1$ Increasing key size may be required for maintaining security level.}
    \end{tablenotes}
\end{threeparttable}
\end{table}

\subsection{Migration Toward PQC}
By formal definition, post-quantum cryptography is cryptography under the assumption that the attacker has a large quantum computer (i.e. the quantum threat discussed in the previous section is realistic), and the central challenge in PQC is to maintain cryptographic usability and flexibility without sacrificing confidence \cite{Bernstein2017}. This requires cryptography to stop relying on the assumed hardness of integer factorization and the discrete logarithmic problem and start using problems that even quantum computers cannot efficiently solve. To date, there have been calls for the Internet to prepare for migration toward Post-Quantum Cryptography (PQC) despite the fact that a large quantum computer has not been built, and the reasons are two-fold: first, secret communications today may be stored and decrypted with QC later and second, pre-quantum public-key cryptosystems like Rivest–Shamir–Adleman~(RSA) and Elliptic Curve Cryptography (ECC) have been embedded within numerous applications and protocols. It is not easy to upgrade software and standards used globally, and such effort must be organized with care. The U.S. National Cybersecurity Center of Excellence's Migration to Post-Quantum Cryptography project \cite{nccoe_2021}, which collaboratively involves many global technology giants, is an example in this direction. The NIST's PQC standardization project \cite{nist-pqc-proj} is another major effort aimed to ``solicit, evaluate, and standardize one or more quantum-resistant public-key cryptographic algorithm''. Started in 2017, the NIST PQC standardization project initially received a total of 69 key exchange mechanism (KEM) and signature scheme proposals, which can be separated into five categories based on their underlying hard problems, as listed in table \ref{table_pqc_taxonomy}. Over the course of 5 years, many have been found unsafe or impractical. Recently, four public-key post-Quantum algorithms have been selected \cite{nist-third-round}, including one key encapsulation mechanism (KEM) and three digital signature schemes. It is expected that at least one additional KEM will also be selected in the near future. A summary of these algorithms is provided in Table \ref{table_nist_algorithms}. Note that two additional stateful hash-based signature schemes are also included in Table \ref{table_nist_algorithms}: the Leighton-Micali Signature (LMS) system and the eXtended Merkle Signature Scheme (XMSS). These two signatures are not candidates of the main NIST PQC standardization project due to the fact that stateful hash-based signatures need to maintain internal states and cannot be implemented using the standard API (i.e. Key Generate, Sign and Verify). As a result, they were approved by the NIST in a separate publication \cite{nist-sp800-208}. 

\begin{table}[]
\centering
\caption{Taxonomy of PQC, their fundamental problems and notable cryptographic schemes submitted to the NIST PQC standardization project.}
\label{table_pqc_taxonomy}
\begin{tabular}{p{1.7cm}|p{4cm}|p{1.7cm}}
\hline\hline
PQC Category                 & Fundamental Problems                                                                                          & Notable Schemes \\ \hline\hline
Lattice-based                & Problems based on structured lattices, such as the (Modular) Learning With Error problem and the NTRU problem & Kyber, Dilithium, Faclon    \\ \hline
Code-based                   & Problems related to decoding linear binary error-correcting code, to which random errors were added.          & McEliece, HQC, BIKE         \\ \hline
Hash-based Signatures        & Combining multiple one-time signature key pairs into a hash tree to generate signatures.                      & SPHINCS+, XMSS              \\ \hline
Isogeny-based                & Problems based on finding isogenies on a large isogeny graph.                                                 & SIKE                        \\ \hline
Multivariate                 & Problems based on finding solutions to large systems of quadratic equations over finite fields.               & Rainbow                     \\ \hline
\end{tabular}
\end{table}

\begin{table}[]
\centering
\caption{List of PQC Algorithms standardized and currently under evaluation by the NIST. }
\label{table_nist_algorithms}
\begin{threeparttable}
    
\begin{tabular}{l|p{1.8cm}|l|l}
\hline\hline
Algorithm        & Foundation             & Type              & NIST Evaluation  \\ \hline\hline
Kyber            & Lattice-based          & KEM               & Recommended      \\ \hline
Dilithium        & Lattice-based          & Signature           & Recommended      \\ \hline
Falcon           & Lattice-based          & Signature           & Alternative      \\ \hline
SPHINCS+         & Stateless Hash-based   & Signature           & Alternative      \\ \hline\hline
HIKE             & Code-based             & KEM               & Under Evaluation \\ \hline
HQC              & Code-based             & KEM               & Under Evaluation \\ \hline
Classic McEliece & Code-based             & KEM               & Under Evaluation \\ \hline\hline
XMSS             & Stateful Hash-based    & Signature           & Recommended$^1$  \\ \hline
LMS              & Stateful Hash-based    & Signature           & Recommended$^1$  \\ \hline
\end{tabular}
\begin{tablenotes}
    \small
    \item $^1$ Recommended in NIST SP 800-208 \cite{nist-sp800-208}, separate from the main NIST PQC standardization project. In general, stateful hash-based cryptography requires careful management of internal states and is harder to implement securely. 
\end{tablenotes}
\end{threeparttable}
\end{table}

\section{Methodology}

We searched digital databases for works that include ``Post-Quantum Cryptography'' and ``Internet of Things'' in the full text. The databases are IEEE Xplore and ACM Digital Library due to their dominance and reputation in the field of computer science. Additionally, we searched the Cryptology ePrint Archive for the newest development. Although works found on the Cryptology ePrint Archive are not peer reviewed, the archive itself has a reputation for its fast processing. Since our work aims to survey the most up-to-date information, this suits our purpose as long as the quality of papers found is critically analyzed. 

The search results were then reviewed and manually filtered. In our scope, a relevant article should:

\begin{itemize}
    \item {be about PQC algorithms and their application on resource-constrained platforms, and}
    \item {evaluate the performance of PQC algorithms, or propose a solution to optimize PQC algorithms for performance, and}
    \item {be published after 2020, as \cite{9363165} has already conducted a survey in 2019. }
\end{itemize}

 A diagram illustrating our filtering process is shown in figure 2. After the initial filtering of search results, 83 publications remained for detailed review and categorization. We found that a total of 34 publications were relevant to the scope of survey: 9 publications are dedicated to the performance evaluation of PQC algorithms on certain resource constrained platforms, whereas the other 25 propose new software designs, hardware designs, or software/hardware co-designs to optimize the performance of PQC algorithms. Tables \ref{table_performance_papers}, \ref{table_software_papers} and \ref{table_hardware_papers} provide lists of literature reviewed, categorized by their primary topics; a pie chart showing the distribution of the research topics is also presented in Figure 3. 

\begin{figure}[]
\centering
\includegraphics[width=3.4 in]{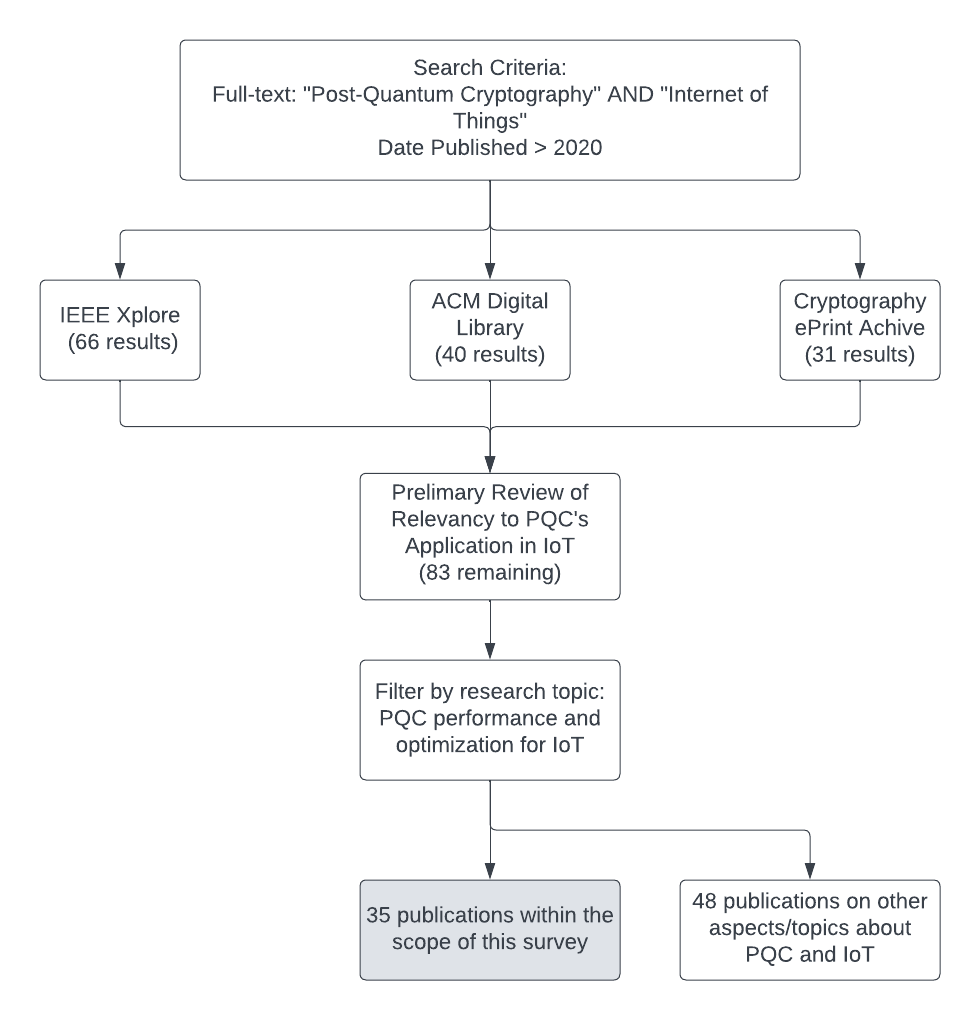}
\caption{Process used to determine publications relevant to our survey's scope. The number of results found by using the search criteria from each database is accurate as of Jan 2023.}
\label{fig_2}
\end{figure}

\begin{figure}[]
\centering
\includegraphics[width=2.5 in]{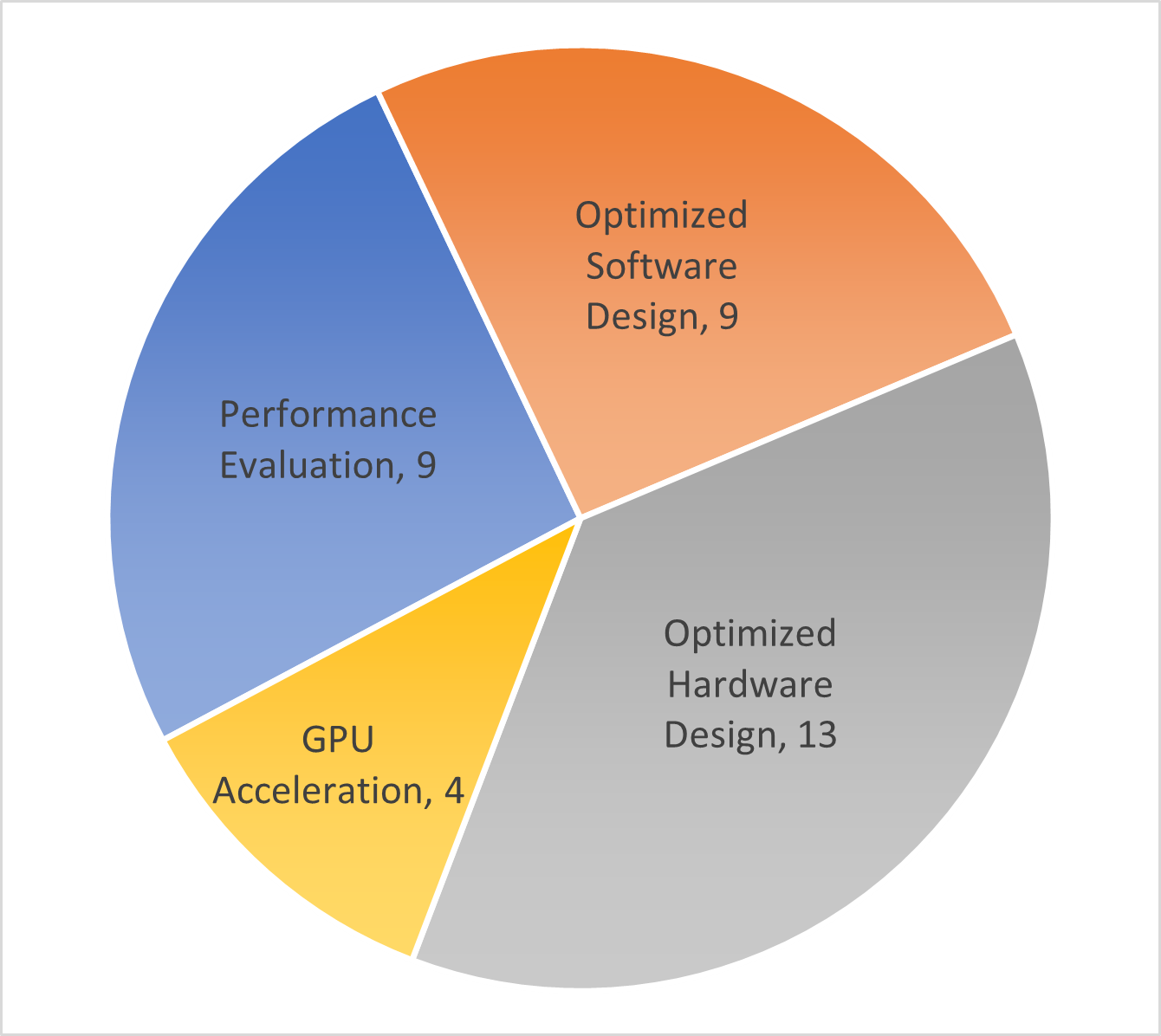}
\caption{Pie chart showing the topic composition of literature surveyed in this paper. }
\label{fig_3}
\end{figure}

\section{PQC Performance Evaluation}

The foremost problem that needs to be addressed when discussing Post-Quantum security for IoT systems is the performance of PQC algorithms, as it was discussed in \cite{8932459} that complex computation may be burdensome for IoT devices. In Table \ref{table_performance_papers}, we summarize recent works on this topic. We notice that the majority of the literature (except for \cite{9679412, 9744589}) include algorithms approved by the NIST (Kyber, Dilithium, Falcon and SPHINCS+) in the evaluation. All works except for one (\cite{10.1145/3507657.3529652}, which focuses only on PQC power consumption) evaluate the speed of PQC algorithms, either in real time or CPU cycles. Such a trend is understandable, as speed/latency is the most intuitive representation of efficiency, and a key requirement for the Next-Generation smart IoT systems \cite{8972389}. However, we must keep in mind that IoT systems often consist of large numbers of small devices, and it is also important to keep other metrics in check: for example, code size (which consumes flash memory), transmission size (which increases bandwidth requirement), RAM usage, and energy/power consumption. As shown in Table \ref{table_performance_papers}, we are yet to find a publication which presents a uniform evaluation framework that considers all those metrics for one or more NIST-selected PQC algorithms operating in resource constrained environments. Lastly, we also note that while \cite{9191980, 9679412, 9744589, 9787987, 10.1145/3507657.3529652} focused on PQC primitives in resource constrained environment, \cite{10.1145/3498361.3538766, 10.1145/3320269.3384725, cryptoeprint:2022/1712, 10.1145/3450268.3453528} implemented the primitives as part of Post-Quantum Transport Layer Security (PQTLS) protocol and evaluated the performance at the transport level. 

Overall, recent works tend to be optimistic in regard to the performance of lattice-based PQC for IoT: despite significant performance overhead, PQC is still shown to be feasible in some reasonably constrained devices. Bürstinghaus-Steinbach et al. \cite{10.1145/3320269.3384725} suggested that the NIST-selected, lattice-based algorithm Kyber runs faster than ECDH by at least one order of magnitude, when both algorithms are executed as primitives on four lightweight devices. However, the code-based SPHINCS+, which was selected by the NIST as an alternative to lattice, demonstrates slower execution than ECDSA. \cite{10.1145/3320269.3384725} further discovered that a Post-Quantum TLS protocol using Kyber-SPHINCS+ cipher suite would introduce a significant burden in terms of latency and memory usage, mainly due to the cost of computing signatures with SPHINCS+. 

The efficiency of Kyber (and lattice-based cryptography in general) is also shown in \cite{9191980, 9787987}. In \cite{9191980}, Mayes demonstrated that the encryption arithmetic of Kyber768 can be run inside a MULTOS Trust Anchor \cite{multos_2022} with just 13 kilobytes of RAM, although the process requires approximately 9.8 seconds even with the aid of a cryptography co-processor. For more powerful devices, \cite{9787987} showed that a Raspberry Pi 4 requires 100-300 microseconds to compute main functions (key generation, encapsulation and decapsulation) of Kyber and SABER, yet lattice-based signatures (Dilithium and Falcon) are more expensive to compute in general. Another study, \cite{10.1145/3507657.3529652}, investigated the power consumption of using Dilithium and Falcon for signature-based authentication. It was found that Dilithium and Falcon's power consumption is in fact on par with RSA-4096, which is commonly used in the pre-quantum era.

A key topic related to the efficiency of PQC in IoT is the Post-Quantum Transport Layer Security (PQTLS). As public-key cryptography is used for key exchange and signature verification in the pre-quantum TLS, their eventual upgrade to PQC should be expected. On a Cortex-M4 micro-controller unit (MCU), \cite{10.1145/3450268.3453528} tested PQKEMs in place of ECDH in TLS 1.2 and measured the overhead, both in terms of latency and energy consumption. It was suggested that the overhead induced by PQC is feasible for IoT, especially with latticed-based KEMs which consumed less energy than ECDH. Furthermore, the authors noted that for lattice-based KEMs, a large portion of latency overhead was caused by larger bandwidth requirements, rather than the actual computation of PQC. As such the potential of hardware acceleration may be limited for PQTLS. A more recent work on this matter is \cite{10.1145/3498361.3538766}, which used the OQS-OpenSSL from Open Quantum Safe to establish TLS 1.3 handshake over MQTT protocol. A key observation noted in this work is that lattice-based KEMs (Kyber, NTRU, NTRU Prime and SABER) achieved faster key exchange than EDCH on both RPi3 and RPi4, despite having to transmit larger packets. However, the same cannot be said for signature authentication as ECDSA outperformed all PQ signature schemes. A key difference between \cite{10.1145/3450268.3453528} and \cite{10.1145/3498361.3538766} is software implementation. \cite{10.1145/3450268.3453528} integrated PQC ciphers within the WolfSSL library, which only supported TLS 1.2 at the time, while the OQS-OpenSSL library used by \cite{10.1145/3450268.3453528} supports TLS 1.3 and was developed specifically for PQC ciphers. Considering the fact that TLS 1.3 is generally faster and superior to TLS 1.2, the information provided by \cite{10.1145/3450268.3453528} might seem more relevant.

If we consider the fact that lattice-based KEMs perform similarly to ECDH yet lattice-based signatures are slower than ECDSA, it would be intuitive to ask if TLS-like authentication can be performed with KEMs only, eschewing the need of computing expensive PQ signatures. Conveniently, KEMTLS has been proposed by Schwabe et al. \cite{10.1145/3372297.3423350}. In KEMTLS protocol, a server uses a certificated long-term KEM public key so that the key exchange can be implicitly authenticated, and no signature has to be generated during the handshake. The certificate used in KEMTLS still requires a CA's PQ signature, but it can be computed offline and only need to be verified once by the client. In \cite{cryptoeprint:2022/1712}, Gonzalez and Wiggers compared KEMTLS and Post-Quantum TLS 1.3 in an embedded setting. It was found that in comparison to PQTLS, KEMTLS can reduce handshake latency by up to 38\%, in addition to achieving less peak memory usage due to the smaller memory footprint of lattice-based KEMs.

\begin{table*}[]
\caption{Summary of literature on the topic of performance evaluation for PQC algorithms} 
\label{table_performance_papers}
\centering
\begin{threeparttable}
    \begin{tabular}{l|l|p{3cm}|p{2.5cm}|p{2.5cm}|p{1.5cm}|llllll}
    \hline
    \multirow{2}{*}{Work}          & \multirow{2}{*}{Year} & \multirow{2}{*}{Algorithm}                                          & \multirow{2}{*}{Implementation}           & \multirow{2}{*}{Hardware}                             & \multirow{2}{*}{Protocols} & \multicolumn{6}{l}{Evaluated Metrics$^1$}                                                                                                                              \\ \cline{7-12} 
                                   &                       &                                                                     &                                           &                                                       &                           & \multicolumn{1}{l|}{S} & \multicolumn{1}{l|}{CS} & \multicolumn{1}{l|}{TS} & \multicolumn{1}{l|}{M} & \multicolumn{1}{l|}{E} & P \\ \hline
    \cite{9191980}                 & 2020                  & Kyber768                                                            & NIST Reference + MULTOS                   & MULTOS IoT Trust-Anchor Board                         & Primitives                & \multicolumn{1}{l|}{$ \checkmark $}     & \multicolumn{1}{l|}{$ \times $}         & \multicolumn{1}{l|}{$ \times $}            & \multicolumn{1}{l|}{$ \times $}      & \multicolumn{1}{l|}{$ \times $}      & $ \times $     \\ \hline
    \cite{10.1145/3320269.3384725} & 2020                  & Kyber, SPHINCS+                                                     & pqm4 + mbedTLS                            & RPi 3, ESP32-PICO-KIT, Fieldbus Option Card, LPC11U68 & Primitives, PQTLS         & \multicolumn{1}{l|}{$ \checkmark $}     & \multicolumn{1}{l|}{$ \checkmark $}         & \multicolumn{1}{l|}{$ \checkmark $}            & \multicolumn{1}{l|}{$ \times $}      & \multicolumn{1}{l|}{$ \times $}      & $ \times $     \\ \hline
    \cite{cryptoeprint:2022/1712}  & 2020                  & Kyber, SABER, NTRU, Dilithium, Falcon                               & pqm4 + WolfSSL + Zephyr RTOS            & STK3701A                                              & KEMTLS, PQTLS             & \multicolumn{1}{l|}{$ \checkmark $}     & \multicolumn{1}{l|}{$ \checkmark $}         & \multicolumn{1}{l|}{$ \checkmark $}            & \multicolumn{1}{l|}{$ \checkmark $}      & \multicolumn{1}{l|}{$ \times $}      & $ \times $     \\ \hline
    \cite{10.1145/3450268.3453528} & 2021                  & McEliece, Kyber, NTRU, NTRU Prime, Saber, FrodoKEM, HQC, BIKE, SIKE & NIST Reference                            & nRF52840                                              & PQTLS                     & \multicolumn{1}{l|}{$ \checkmark $}     & \multicolumn{1}{l|}{$ \times $}         & \multicolumn{1}{l|}{$ \times $}            & \multicolumn{1}{l|}{$ \times $}      & \multicolumn{1}{l|}{$ \checkmark $}      & $ \times $     \\ \hline
    \cite{9679412}                 & 2021                  & Mersenne-756839                                                     & Not specified $^2$                            & ESP8266                                               & Primitives                & \multicolumn{1}{l|}{$ \checkmark $}     & \multicolumn{1}{l|}{$ \times $}         & \multicolumn{1}{l|}{$ \checkmark $}            & \multicolumn{1}{l|}{$ \times $}      & \multicolumn{1}{l|}{$ \checkmark $}      & $ \times $     \\ \hline
    \cite{9744589}                 & 2022                  & NTRU                                                                & pqm4                                      & EFR32MG12                                             & Primitives                & \multicolumn{1}{l|}{$ \checkmark $}     & \multicolumn{1}{l|}{$ \times $}         & \multicolumn{1}{l|}{$ \times $}            & \multicolumn{1}{l|}{$ \checkmark $}      & \multicolumn{1}{l|}{$ \checkmark $}      & $ \checkmark $     \\ \hline
    \cite{9787987}                 & 2022                  & Kyber, Saber, Dilithium, Falcon                                     & liboqs                                    & RPi 4                                                 & Primitives                & \multicolumn{1}{l|}{$ \checkmark $}     & \multicolumn{1}{l|}{$ \times $}         & \multicolumn{1}{l|}{$ \times $}            & \multicolumn{1}{l|}{$ \times $}      & \multicolumn{1}{l|}{$ \times $}      & $ \times $     \\ \hline
    \cite{10.1145/3507657.3529652} & 2022                  & Dilithium, Falcon                                                   & Not specified $^2$                            & RPi 3 \& 4; Laptop and PC                             & Primitives                & \multicolumn{1}{l|}{$ \times $}     & \multicolumn{1}{l|}{$ \times $}         & \multicolumn{1}{l|}{$ \times $}            & \multicolumn{1}{l|}{$ \times $}      & \multicolumn{1}{l|}{$ \times $}      & $ \checkmark $     \\ \hline
    \cite{10.1145/3498361.3538766} & 2022                  & BIKE, HQC, FrodoKEM, Kyber, NTRU, NTRU Prime, SABER, SIKE           & NIST Reference + existing application & RPi 3 \& 4                                            & PQTLS                     & \multicolumn{1}{l|}{$ \checkmark $}     & \multicolumn{1}{l|}{$ \times $}         & \multicolumn{1}{l|}{$ \times $}            & \multicolumn{1}{l|}{$ \times $}      & \multicolumn{1}{l|}{$ \times $}      & $ \times $     \\ \hline
    \end{tabular}
    \begin{tablenotes}
      \small
      \item $^1$ Acronyms used in the header: S - Speed; CS - Code Size; TS - Transmission Size; M - Memory; E - Energy; P - Power.
      \item $^2$ These works did not state explicitly which implementation was used for evaluation, nor did the authors provide the source code. It is likely that the C reference implementation submitted to the NIST was used, as both works recognized the algorithms as NIST PQC standardization candidates. 
    \end{tablenotes}
\end{threeparttable}
\end{table*}

\section{Software Optimization for PQC}



As shown in Table \ref{table_software_papers}, recent works on software optimization for PQC focused heavily on lattice-based cryptography (LBC), as 7 out of a total 9 papers described means to optimize lattice-based KEM and signatures. However, papers published in 2020 did not seem to focus on a specific type of PQC: ways to optimize hash-based (XMSS), lattice-based (NTRU Primes, Saber) and a cipher based on zero-knowledge proofs, namely Picnic, were discussed. However, after 2021 there seems to be a shift in direction, as all 5 papers focused on optimizing LBC algorithms, including Dilithium and Falcon which were officially selected by the NIST for standardization. 

In regard to optimization goals, we note that proposals for PQC software optimization aim all to improve operation speed (latency) and memory usage. In general, the quest for optimization is a \emph{trade-off problem} between speed and memory usage. For example, a relatively straight-forward optimization technique for reducing minimum memory requirement is to generate data required for computation (i.e. coefficient matrices of large polynomial systems) only when needed, instead of storing all data in the memory space simultaneously. This technique was used in \cite{9217806, cryptoeprint:2022/323, 9682594} to drastically reduce memory usage at the cost of extra computation time. For LBC, the most computational intensive task is polynomial multiplication: as practical LBC algorithms need to compute multiplications of large and high-degree polynomials, the optimization of this operation is rewarding. In a ``school book'' implementation, polynomial multiplication can be done by multiplying each pair of coefficients, which has a time complexity of \(O(n^2)\), where \(n\) is the degree of two polynomials being multiplied together. Alternatively, an algorithm which operates on the similar principles of Fast Fourier Transform (FFT) - called Number Theoretic Transform (NTT) - can reduce this complexity to the order of \(O(n)\log n\) \cite{fft-algorithm}. Broadly speaking, like FFT and inverse-FFT, NTT and inverse-NTT transforms polynomials to and back from points representation using values evaluated at roots of unity. Yet unlike FFT which uses n-th complex roots of unity (i.e. complex number \(z\) where \(z^n = 1\)), NTT uses roots of unity in a finite field (i.e. modular arithmetic over a prime number). Some papers demonstrate that it is possible to modify the standard NTT algorithm for further enhancement in performance: for example, \cite{9217806} notices that some polynomials used in Dilithium have small coefficients and proposes a parallel algorithm which runs faster with ``small'' polynomials than the standard NTT. \cite{cryptoeprint:2022/323}, on the other hand, suggests that by performing NTT over a smaller prime number, some polynomial coefficient values can be stored as 16-bit value instead of 32-bit. However, this deviates from the standard specification of Dilithium, which raises questions about whether or not such modification affects the scheme's security strength - unfortunately, such issues were not fully discussed. Whilst other works optimize the algorithms by making trade-offs between speed and memory usage, Alkim et al. \cite{cryptoeprint:2020/1216} claimed to achieve both. The authors present three means to optimize the NTRU Prime KEM and claim that one is optimized for speed using Good's trick \cite{good_1951}, whilst the other two use mixed-radix NTT multiplication and are optimized for memory usage. Despite the fact that these methods focus on different metrics, all of them seem to outperform the standard pqm4 implementation - which is already a PQC library optimized for the Cortex-M4 platforms \cite{PQM4} - in both speed and memory. The only downside appears to be the fact that those optimizations only support a limited set of parameters used in NTRU Prime. Nevertheless, we should note that NTRU Prime is no longer a KEM being considered for NIST standardization, but this work does raise the question of whether a joint speed-and-memory optimization is possible for other LBC algorithms.

On the other hand, for the stateful hash-based signature (HBS) XMSS, \cite{cryptoeprint:2020/898} discusses an interesting technique for maximizing speed for signature verification on lightweight devices. One key characteristic of stateful HBS is that the private key used by the signer consists of a large set of one-time signature (OTS) keys, and an OTS key should never be reused \cite{nist-sp800-208}. Additionally, the verification of XMSS signatures is not a constant-time operation, as the time to verify signatures (of the same message) generated by different OTS varies for each OTS. Therefore the authors propose an altered XMSS algorithm, in which the signer uses a probabilistic method to search for signatures that are faster to verify. The addition of this ``search'' process, in practice, seems to drastically increase the signing time: on a general purpose CPU, the increase in signing time may reach one minute, and this does not seem feasible for any real-time network application that requires frequent signature-based authentication - for example, a TLS-like protocol for server authentication.

Instead of tweaking the internal workings of cryptographic algorithms, a more conservative approach might be optimizing lower level software and catering to the needs of cryptographic computations. In \cite{9762260}, Kim et al. took advantage of ARMv8's Neon technology and parallelized the overall process of FFT/NTT-based polynomial multiplication and reported 15.1\%, 16.4\% and 65.4\% speed improvement in key generation, signing and verification, respectively. Considering the fact that Falcon by design is a signature scheme that features exceptionally fast verification but slower key generation and signing, the drastic improvement in signature verification may further push Falcon as the preferred scheme in scenarios where a large number of client-side lightweight IoT devices are frequently tasked to verify server signatures. On the other hand, for the RISC-V architecture, \cite{cryptoeprint:2021/1117} implemented faster multiplication methods and designed a new processor instruction set extension dedicated to polynomial multiplication. However, although  \cite{cryptoeprint:2021/1117} claims improvement in performance, no comparison with the existing state-of-the-art method (i.e. traditional NTT) is provided. As such, we cannot practically evaluate this implementation's potential improvement on a post-quantum KEM or signature scheme. 

\begin{table*}[]
\caption{Summary of literature on the topic of software optimization for PQC algorithms on resource-constrained devices}
\label{table_software_papers}
\centering
\begin{threeparttable}
    \begin{tabular}{l|l|l|l|p{7cm}|ll}
    \hline\hline
    \multirow{2}{*}{Work}          & \multirow{2}{*}{Year} & \multirow{2}{*}{Algorithm}  & \multirow{2}{*}{Target Hardware} & \multirow{2}{*}{Optimization Technique}                                                                                                   & \multicolumn{2}{c}{Metrics Considered$^1$} \\ \cline{6-7} 
                                   &                       &                            &               &                                                                                                                            & \multicolumn{1}{l|}{Speed}  & Memory  \\ \hline\hline
    \cite{cryptoeprint:2020/898}   & 2020                  & XMSS  & Cortex-M4          & Using probabilistic method to generate signatures that are easier to verify, at the cost of longer signing time.                           & \multicolumn{1}{l|}{+}        & -       \\ \hline
    \cite{cryptoeprint:2020/1216}  & 2020                  & NTRU Prime & Cortex-M4     & Good's trick for improving speed and mixed radix NTT for improving memory usage.                                                          & \multicolumn{1}{l|}{+}        & +       \\ \hline
    \cite{9217806}                 & 2020                  & Picnic   & -               & Generating data on demand; avoid caching; streaming signature output.                                                                     & \multicolumn{1}{l|}{-}        & +       \\ \hline
    \cite{cryptoeprint:2020/1303}  & 2020                  & Saber    & -               & Using alternative algorithms to convert polynomial multiplication into large integer multiplication suitable for acceleration with CCoPs. & \multicolumn{1}{l|}{+}        & ?       \\ \hline
    \cite{cryptoeprint:2021/1117}  & 2021                  & General LBC & RISC-V       & Efficient instruction sets for 32-bit RISC-V processors suitable for polynomial multiplication.                                           & \multicolumn{1}{l|}{+}        & ?       \\ \hline
    \cite{cryptoeprint:2022/323}   & 2022                  & Dilithium & Cortex-M4      & Streaming and compressing data; alternative algorithms for polynomial multiplication.                                                     & \multicolumn{1}{l|}{-}        & +       \\ \hline
    \cite{10.1145/3564625.3564629} & 2022                  & Dilithium & Cortex-A series & Using parallel small polynomial multiplication to accelerate Sign and Verify operations for Dilithium.                                    & \multicolumn{1}{l|}{+}        & -       \\ \hline
    \cite{9762260}                 & 2022                  & Falcon & Cortex-A series    & Utilizing NEON engine and vector instructions to parallelize FFT/NTT-based polynomial multiplication.                                     & \multicolumn{1}{l|}{+}        & ?       \\ \hline
    \cite{9682594}                 & 2022                  & Saber & RISC-V              & Finding optimized NTT parameters for polynomial multiplication; generating data on demand.                                                & \multicolumn{1}{l|}{-}        & +       \\ \hline
    \end{tabular}

    \begin{tablenotes}
        \small
        \item{$^1$ Symbols are used to indicate whether a proposed implementation intends to improve (+), sacrifice (-) a metric. Question mark (?) indicates that a metric is not included in the evaluation. }
    \end{tablenotes}
\end{threeparttable}

\end{table*}

\section{Hardware Optimization for PQC}

Another practical way to optimize cryptography for performance is implementing cryptographic functions in hardware, such as using Field Programmable Gate Arrays (FPGAs) or Application Specific Integrated Circuits (ASICs). In general, hardware implementation is faster than software implementation of the same function, yet the benefit in this case is not only execution speed: the need for cryptographic functions to compete with application logic for memory space can also be eliminated by moving those functions to separate hardware, and this may be crucial for memory constrained IoT devices. However, such perks come with a cost: the introduction of extra hardware would also cost extra energy, and this problem is more prominent for FPGAs than ASICs \cite{s22197496}. On the other side, ASICs sacrifice flexibility (i.e. unlike FPGAs, ASICs cannot be re-programmed once produced) for performance, and generally have higher development cost along with longer time-to-market \cite{9001747}. 

Before the discussion of specific works in Table \ref{table_hardware_papers}, we should keep in mind that a common issue with hardware optimization is the lack of standards. For software, NIST's reference implementations in C and the pqm4 library are used as the \emph{de facto} standard. For hardware optimization, many work report improvement in some aspects over a previous similar hardware design, but as a common baseline does not seem to exist, cross-comparing is rather difficult. Therefore, on the right-hand-side of Table \ref{table_hardware_papers} where we indicate whether each optimization ``improves'' or ``sacrifices'' a metric, the indications are not relative to a global or absolute baseline, but a summary of comparison made to certain previous works selected by the authors.

Nevertheless, we surveyed papers listed in Table \ref{table_hardware_papers}. We notice that unlike optimized software implementations, recent works on hardware PQC are less focused on NIST PQC candidates. 4 out of 13 surveyed papers propose optimization for the Binary Ring Learning With Error (Bin-RLWE) \cite{10.1145/2899007.2899011}, which seems to be a popular cipher among hardware designers \cite{9211858, 9717291, 9737700, 10.1145/3569457}. One possible reason for this is that Bin-RLWE is by design lightweight and implementation-friendly, as the binary error values can be efficiently generated in hardware \cite{8342207}. Yet the Bin-RLWE was not submitted as a NIST standardization candidate. As such, the likelihood of it being widely accepted is not as high as other lattice-based schemes, despite its inherent suitability for hardware. Overall, we note that the most promising hardware optimization for Bin-RLWE is \cite{10.1145/3569457}, where He et al. demonstrate that using the combination of algorithmic derivation and parallel computation, a high area-delay product (ADP) can be achieved by improving area and speed simultaneously, rather than sacrificing one for another. 

Kim et al. \cite{9911531} is another recent work that we found significant for the current PQC optimization landscape. This work uses a look-up-table-based approach for NTT modular multiplication, and a real-time processing for polynomial sampling. These optimization techniques were applied on ASIC and an overall improvement in all metrics (speed, area, memory and energy) were reported. As reported in the paper, this implementation can be used for any LBC where the modulus parameter ($q$) is smaller than $2^{24}$ - which both Dilithium ($q = 8,380,417$) and Kyber ($q = 7,681$) satisfy. Therefore, it would be interesting to investigate how well this design (and ASIC crypto-accelerators in general) applies to the NIST-selected algorithms in practice in future studies.  

On specific standardized LBC ciphers, one recent work \cite{cryptoeprint:2022/217} provided a hardware implementation for the NIST-selected lattice signatures, Dilithium and Falcon. In this work, Beckwith et al. discuss the challenges of implementing NTT for polynomial multiplication, as well as performance bottle-necking caused by the need of large pseudo-random data for polynomial sampling in Dilithium. Addressing these challenges, the authors present an optimized Dilithium implementation in FPGA, which claims the best latency to date. For Dilithium-II on a Xilinx Artix-7 FPGA board running at 256 MHz, their implementation reported 19, 27 and 117 microseconds on average for key generation, verifying and signing operations, respectively. However, the issue with energy was left not discussed: for a small IoT device, the addition of an FPGA is likely to increase the device's power consumption. Should similar hardware implementation be used in a practical design, a well-understood energy profile is crucial. The work in~\cite{cryptoeprint:2022/217} provides the first hardware implementation of Falcon (albeit only the verify function) and achieved a latency as low as 16.8 microseconds on an Artix-7 at 142 MHz. Again, the energy-efficiency of this approach comes into question, considering the FPGA's higher energy demand and the lack of energy-wise evaluation in this work. 

\cite{9460703} is another work with notable significance as it features an area-efficient hardware implementation of SPHINCS+, which is a stateless hash-based signature also selected by the NIST (albeit not a primary recommendation, as shown in many works such as \cite{10.1145/3320269.3384725}, its computation is not as efficient as LBC). Instead of speed, this work proposes an FPGA implementation that maximizes area-efficiency yet still maintains acceptable latency. The authors show that this implementation of SPHINCS+, on a Xilinx-XZU3EG FPGA running at ~150MHz frequency can achieve comparable speed to software implementation running on a 3.6GHz Intel-i7 CPU: 2.02 milliseconds for key generation, 64.34 milliseconds for signing, and 2.51 milliseconds for verifying. However, we should note that such results are three orders of magnitude higher than Dilithium and Falcon, the operations of which are measured in tens of microseconds. Naturally, this raises the question if hash-based cryptography is suitable for resource-constrained devices, if LBC signatures can perform ~100 times faster than HBC on similar devices (27 microseconds for Dilithium and 16.8 microseconds for Falcon, as reported in \cite{cryptoeprint:2022/217}.  

\begin{table*}[]
\caption{Summary of literature on the topic of hardware optimization for PQC algorithms on FPGA and ASIC}
\label{table_hardware_papers}
\centering
\begin{threeparttable}

\begin{tabular}{l|l|p{1.3cm}|l|p{7.5cm}|llll}
\hline\hline
\multirow{2}{*}{Work}         & \multirow{2}{*}{Year} & \multirow{2}{*}{Algorithm} & \multirow{2}{*}{Hardware} & \multirow{2}{*}{Optimization Technique}                                                                                                                                  & \multicolumn{4}{l|}{Metrics Considered$^1$}                                                             \\ \cline{6-9} 
                              &                       &                            &                           &                                                                                                                                                                          & \multicolumn{1}{l|}{Speed}    & \multicolumn{1}{l|}{Area}     & \multicolumn{1}{l|}{Power} & Memory     \\ \hline\hline
\cite{9211858}                & 2020                  & Bin-RLWE                   & FPGA                      & Novel masked implementation method against differencial power attack without using a pseudo random number generator.                                                     & \multicolumn{1}{l|}{+}        & \multicolumn{1}{l|}{+}        & \multicolumn{1}{l|}{?}     & $ \times $ \\ \hline
\cite{9180839}                & 2020                  & RLWE                       & ASIC                      & Using novel Approximate Modular Multiplier design for faster RLWE.                                                                                                       & \multicolumn{1}{l|}{+}        & \multicolumn{1}{l|}{+}        & \multicolumn{1}{l|}{+}     & ?          \\ \hline
\cite{8877876}                & 2020                  & LEDAkem                    & FPGA                      & Optimized block partitioning and lazy accumulation techniques.                                                                                                           & \multicolumn{1}{l|}{?}        & \multicolumn{1}{l|}{+}        & \multicolumn{1}{l|}{?}     & ?          \\ \hline
\cite{9089230}                & 2020                  & LEDAsig                    & FPGA                      & Compact design with novel technique for quasi-cyclic block rotation by using \emph{read-first} feature of BRAMs.                                                         & \multicolumn{1}{l|}{-}        & \multicolumn{1}{l|}{+}        & \multicolumn{1}{l|}{?}     & -          \\ \hline
\cite{9460703}                & 2021                  & SPHINCS+                   & FPGA                      & Area-efficient implementation using a data path around a sequential SHA-256 core, and a control path with nested Finite State Machines.                                  & \multicolumn{1}{l|}{-}        & \multicolumn{1}{l|}{+}        & \multicolumn{1}{l|}{?}     & -          \\ \hline
\cite{9717291}                & 2022                  & Bin-RLWE                   & FPGA                      & Efficient vector multiplication with simple shift-and-add algorithm over a finite ring.                                                                                  & \multicolumn{1}{l|}{+}        & \multicolumn{1}{l|}{-}        & \multicolumn{1}{l|}{?}     & +          \\ \hline
\cite{9911531}                & 2022                  & General LBC                & ASIC                      & Look-up-table-based modular multiplication and real-time processing for polynomial sampling.                                                                             & \multicolumn{1}{l|}{+}        & \multicolumn{1}{l|}{+}        & \multicolumn{1}{l|}{+}     & +          \\ \hline
\cite{9712707}                & 2022                  & General LBC                & ASIC                      & Accelerating NTT by parallelizing calculations with multiple arithmetic units in circuit design.                                                                         & \multicolumn{1}{l|}{+}        & \multicolumn{1}{l|}{-}        & \multicolumn{1}{l|}{?}     & ?          \\ \hline
\cite{9559627}                & 2022                  & XMSS                       & FPGA                      & Hardware implementation with optimized node traversal for the L-tree and Merkel tree; utilizing the Buchmann-Dahmen-Schneider algorithm for faster signature generation. & \multicolumn{1}{l|}{+}        & \multicolumn{1}{l|}{?}        & \multicolumn{1}{l|}{?}     & $ \times $ \\ \hline
\cite{9869764}                & 2022                  & XMSS                       & FPGA                      & Two designs that use multiple hash cores for concurrent computation.                                                                                                     & \multicolumn{1}{l|}{+}        & \multicolumn{1}{l|}{-}        & \multicolumn{1}{l|}{+}     & $ \times $ \\ \hline
\cite{9737700}                & 2022                  & Bin-RLWE                   & FPGA                      & Novel algorithmic operations for polynomial multiplication and addition.                                                                                                 & \multicolumn{1}{l|}{+/- $^2$} & \multicolumn{1}{l|}{-/+ $^2$} & \multicolumn{1}{l|}{?}     & $ \times $ \\ \hline
\cite{10.1145/3569457}        & 2022                  & Bin-RLWE                   & FPGA                      & Novel implementation-oriented algorithmic derivation for achieving low implementation complexity.                                                                        & \multicolumn{1}{l|}{+}        & \multicolumn{1}{l|}{+}        & \multicolumn{1}{l|}{?}     & $ \times $ \\ \hline
\cite{cryptoeprint:2022/217}  & 2022                  & Dilithium and Falcon       & FPGA                      & Overall high performance design for Dilithium, and first hardware implementation for Falcon-verify.                                                                      & \multicolumn{1}{l|}{+}        & \multicolumn{1}{l|}{+}        & \multicolumn{1}{l|}{?}     & -          \\ \hline
\end{tabular}

    \begin{tablenotes}
        \small
        \item{$^1$ Symbols are used to indicate whether a proposed implementation improves (+), sacrifices (-) or omits (?) a metric. Additionally, ``$\times$'' in the memory column indicates that Block Random Access Memory (BRAM) is not used in the corresponding FPGA design.}
        \item{$^2$ Two different designs are proposed: one optimizes area-complexity against speed and the other does the opposite. However, both designs achieve higher area-delay product against previous works.} 
    \end{tablenotes}
\end{threeparttable}
\end{table*}

\section{Graphic Processor Unit (GPU) Acceleration for PQC}

Another category of papers we surveyed  is on the topic of GPU acceleration for PQC algorithms. GPU has long been used for cryptographic computation due to the advantage offered by its inherent parallel architecture \cite{10.1007/978-3-540-85053-3_6}. Aside from its original purpose (graphic rendering), a major use case of GPU computation is the mining of crypto-currencies \cite{su14148708}, in which a GPU's large number of parallel cores can be used to reach very high throughput compared to general-purpose CPUs. Similarly, the advantage offered by GPU computation can be leveraged for PQC - especially for LBC, as the key to accelerating LBC is parallelizing expensive polynomial multiplication (as per earlier discussion). Among works we surveyed four papers are dedicated to GPU acceleration for LBC \cite{9860310, 9667701, 9715141, cryptoeprint:2021/1389} as shown in Table \ref{table_gpu_papers}. A commonality among these papers is that their ``optimization'' is purely aimed for computation speed/throughput without consideration for other metrics - after all, the GPUs used in their implementation are powerful hardware with high energy demand, which can execute hundreds of thousands LBC key encapsulation and decapsulation per second. For example, \cite{9667701} reports that with an optimized GPU algorithm for SABER, a single GPU can reach a key generation throughput that is equivalent to 64 CPU cores, and \cite{cryptoeprint:2021/1389} shows that a NVIDIA RTX3080 GPU can process 124,418 key exchanges per second using SABER, whereas a Intel Core i9-10900K can only process 29,836 per second. 

For smaller devices, Lee et al. \cite{9511134} show that the same technique is also feasible on a Jetson Nano, which is an embedded device with a Cortex-A53 processor and a small 128-core GPU. It is shown that by utilizing GPU acceleration on the Jetson Nano, Kyber-512 can achieve a throughput performance of 1,345 key encapsulations per second and 918 key decapsulations per second. Although \cite{9511134} does not mention how this result compares to using CPU, it is possible to cross-check it against \cite{9787987}, which reports that on a Raspberry Pi 3 (which uses the same Cortex-A53 processor), the average execution time for Kyber-512 is 1.1 ms - or 909 operations for second - for both encapsulation and decapsulation. As the same paper reports that Kyber performs more than 9 times faster with GPU acceleration (RTX2060) than using CPU alone (Intel i9-9700F), it's reasonable to suspect that for lightweight devices, accelerating PQC computation with small GPUs may not be as rewarding. However, more data is needed on this topic to obtain a definitive answer. 

Regardless of the size, we should note that the usage of GPU is not common in resource-constrained IoT nodes. An argument repeatedly used by Lee et al. \cite{9860310, 9715141, cryptoeprint:2021/1389} is that GPUs may be used in gateway devices to provide ``Encryption-as-a-Service'' for IoT nodes in proximity. Whilst such novelty may prove its value in a sophisticated IoT network, it does not directly contribute to solving the resource-constraint problem at the node layer: unless the node devices are also capable of dynamic post-quantum key-exchange and authentication, this arrangement does not improve the security of node-to-gateway communication. As such, it is reasonable to argue that the application of GPU acceleration for PQC on resource-constrained platforms may be limited.

\begin{table*}[]
\centering
\caption{Summary of literature on the topic of GPU acceleration for PQC algorithms}
\label{table_gpu_papers}
\begin{threeparttable}
\begin{tabular}{l|l|p{1.5cm}|p{3cm}|p{7cm}}
\hline\hline
Work                          & Year & Algorithm           & GPU                          & Optimization Technique                                                                                       \\ \hline \hline
\cite{9511134}                & 2021 & Kyber               & NVIDIA RTX2060; 128-core Maxwell GPU & Accelerating Kyber KEM's encapsulation and decapsulation using a fine-grain parallel GPU implementation.       \\ \hline
\cite{9667701}                & 2021 & Saber               & NVIDIA A100                  & Accelerating Saber key generation on GPU with Physical Unclonable Functions (PUFs).                            \\ \hline
\cite{9715141}                & 2022 & NTRU, LAC, FrodoKEM & NVIDIA RTX2060               & Accelerating LBC polynomial multiplication with NVIDIA Tensor Core Technology.                                   \\ \hline
\cite{cryptoeprint:2021/1389} & 2022 & FrodoKEM, Saber     & NVIDIA RTX 3080, V100 and T4 & Accelerating LBC polynomial multiplication with dot-product instructions available on NVIDIA GPUs. \\ \hline
\end{tabular}
\end{threeparttable}
\end{table*}

\section{Critical Reflection and Future Directions}

In this paper, we surveyed 35 recent publications on the performance and optimization of PQC for IoT applications. First we reviewed papers which evaluate the performance of PQC - both as cryptographic primitive functions and as parts of TLS-like protocols - in resource-constrained devices. Then we examined optimized software and hardware PQC implementations, which may extend the protection of PQC to lightweight systems. Finally, we reviewed recent attempts GPU acceleration for PQC, which may be possible on the IoT gateway layer. 

On performance evaluation, we note that lattice-based KEMs seem to be feasible in reasonably constrained devices, and some KEMs were reported to outperform the current state-of-the-art key exchange algorithms in terms of speed. However, post-quantum signature schemes appear to have slower computation speed and heavier memory footprint than their pre-quantum counterparts. As a result, the KEMTLS - an alternative TLS-like protocol which does not need on-the-fly signature generation and uses long-term KEM public keys to implicitly authenticate users - seems to be a convenient arrangement for resource-constrained IoT networks. Furthermore, we realize that recent works all focus on a limited selection of PQC algorithms, and metrics used in these works are not uniform. We suggest that a comprehensive PQC evaluation framework should be proposed specifically for resource-constrained platforms so that IoT applications in the post-quantum era can be designed efficiently and safely. 

On optimized software designs, recent development in this area mainly focuses on the trade-off between speed and memory usage. We reviewed the implementations and summarize that the recent proposals generally fall into one of the three categories: lazy initiation (for reducing peak memory usage at the cost of speed), parallelization (for accelerating speed at the cost of memory usage and design complexity) and alternative polynomial multiplication for lattice-based algorithms. In general, there is no "free lunch" in this regard and the feasibility of those trade-off designs depends on the application context. We also note that some proposals were made to optimize PQC algorithms that are no longer being considered for standardization by the NIST, making them less relevant today. As a result, we believe that exploration in this direction should focus on algorithms already selected by the NIST to maximize research contribution. 

On hardware implementations, we found that recently proposed schemes are no longer trade-offs, and it is possible to improve several metrics simultaneously by utilizing efficient designs. We observed that many proposals report improvement in both speed and area complexity, but power consumption seems to be a metric on which less attention has been paid. Overall, the performance improvements achieved by ASIC designs seem to be more significant than that of FPGA designs, but it should be kept in mind that ASICs lack flexibility and generally have longer time-to-market. Additionally, we note the lack of ``baseline'' for hardware makes cross-comparison difficult, especially when specific algorithms have received significantly less attention in this regard. The lack of attention on hardware implementation for the NIST-selected candidates might be due to the fact that an alternative LBC algorithm - the more lightweight, hardware-friendly Bin-RLWE - is simpler to implement. However, we argue that because the Bin-RLWE is not a NIST PQC candidate, the likelihood of it receiving global recognition is low and more attention should be paid to the primary NIST recommendations (i.e. Kyber and Dilithium) in the future. 

Finally, GPU acceleration for PQC is a less explored field to date despite the long history of using GPU for (pre-quantum) cryptographic computation. We note that as far as high-end GPUs are concerned, the benefit of GPU acceleration is significant: it is possible to reach a throughput of hundreds of thousands of key exchanges per second on a single GPU. However, it seems that smaller GPUs found on SoCs do not offer the same advantage against lightweight Cortex-A series CPUs. This may potentially make GPU acceleration less suitable for resource-constrained IoT systems, as high-end GPUs often have high power demand and the possibility of integrating them with IoT devices on the node layer is slim.

Overall, we observe that existing literature on the performance of PQC in resource-constrained IoT applications is providing interesting clues, but is also far from comprehensive. On the other hand, as the NIST's PQC standardization project is expected to publish the complete standard by 2024 \cite{nist-third-round}, we believe that researchers should seek alignment with the standardization efforts. In other words, we suggest that future research should focus on the NIST-approved KEMs and signatures, rather than novel cryptosystems that are less likely to receive global adaptation. 

\section{Conclusion}

Quantum computing poses great threat to public key cryptosystems that are widely used today, but the computation of PQC may be heavyweight for IoT applications, which typically employs a large number of lightweight and resource-constrained devices. In this paper, 35 recent publications are surveyed and the latest developments on performance evaluation, software/hardware optimization, and GPU acceleration for PQC are discussed in the context of IoT. By tabulating and analyzing their findings, we summarize and present key discoveries in this area. We further conclude that whilst some PQC schemes are feasible for reasonably lightweight IoT devices, proposals for their optimization seem to lack focus and standardization. As the NIST is set to publish the first standard for PQC in 2024, we suggest future research to seek coordination, in order to ensure an efficient and safe migration toward IoT for the post-quantum era.

\bibliographystyle{ieeetr} 
\bibliography{bibtex/bib/IEEEabrv, ref}

\begin{thebibliography}{10}

\bibitem{Hasan2022}
M.~Hasan, ``State of iot 2022: Number of connected iot devices growing 18\% to
  14.4 billion globally.''
  https://iot-analytics.com/number-connected-iot-devices/, Jun 2022.
\newblock Accessed: 2023-3-24.

\bibitem{Shor1994}
P.~W. Shor, ``Algorithms for quantum computation: discrete logarithms and
  factoring,'' in {\em Proceedings 35th annual symposium on foundations of
  computer science}, pp.~124--134, Ieee, 1994.

\bibitem{Grover1996}
L.~K. Grover, ``A fast quantum mechanical algorithm for database search,'' in
  {\em Proceedings of the twenty-eighth annual ACM symposium on Theory of
  computing}, pp.~212--219, 1996.

\bibitem{Malina2021}
L.~Malina, P.~Dzurenda, S.~Ricci, J.~Hajny, G.~Srivastava, R.~Matulevičius,
  A.-A.~O. Affia, M.~Laurent, N.~H. Sultan, and Q.~Tang, ``Post-quantum era
  privacy protection for intelligent infrastructures,'' {\em IEEE Access},
  vol.~9, pp.~36038--36077, 2021.

\bibitem{8932459}
T.~M. Fernández-Caramés, ``From pre-quantum to post-quantum iot security: A
  survey on quantum-resistant cryptosystems for the internet of things,'' {\em
  IEEE Internet of Things Journal}, vol.~7, pp.~6457--6480, July 2020.

\bibitem{nist-pqc-proj}
NIST, ``Post-quantum cryptography.''
  https://csrc.nist.gov/projects/post-quantum-cryptography, Jan 2017.
\newblock Accessed 2023-3-24.

\bibitem{PQM4}
M.~J. Kannwischer, R.~Petri, J.~Rijneveld, P.~Schwabe, and K.~Stoffelen,
  ``{PQM4}: Post-quantum crypto library for the {ARM} {Cortex-M4}.''

\bibitem{9363165}
L.~Malina, P.~Dzurenda, S.~Ricci, J.~Hajny, G.~Srivastava, R.~Matulevičius,
  A.-A.~O. Affia, M.~Laurent, N.~H. Sultan, and Q.~Tang, ``Post-quantum era
  privacy protection for intelligent infrastructures,'' {\em IEEE Access},
  vol.~9, pp.~36038--36077, 2021.

\bibitem{nist-third-round}
G.~Alagic, D.~Apon, D.~Cooper, Q.~Dang, T.~Dang, J.~Kelsey, J.~Lichtinger,
  Y.-K. Liu, C.~Miller, D.~Moody, R.~Peralta, R.~Perlner, A.~Robinson, and
  D.~Smith-Tone, ``Status report on the third round of the nist post-quantum
  cryptography standardization process.''
  https://nvlpubs.nist.gov/nistpubs/ir/2022/NIST.IR.8413-upd1.pdf, Jul 2022.

\bibitem{seyhan2021}
K.~Seyhan, T.~N. Nguyen, S.~Akleylek, and K.~Cengiz, ``Lattice-based
  cryptosystems for the security of resource-constrained iot devices in
  post-quantum world: A survey,'' {\em Cluster Computing}, vol.~25, no.~3,
  p.~1729–1748, 2021.

\bibitem{ieft-rc-classification}
C.~Bormann, M.~Ersue, and A.Keranen, ``Rfc 7228: Terminology for
  constrained-node networks.'' https://www.rfc-editor.org/info/rfc7228, 2014.
\newblock Accessed: 2023-3-24.

\bibitem{zolotova2015}
I.~Zolotov{\'a}, M.~Bundzel, and T.~Lojka, ``Industry iot gateway for cloud
  connectivity,'' in {\em Advances in Production Management Systems: Innovative
  Production Management Towards Sustainable Growth: IFIP WG 5.7 International
  Conference, APMS 2015, Tokyo, Japan, September 7-9, 2015, Proceedings, Part
  II 0}, pp.~59--66, Springer, 2015.

\bibitem{urien-lwig-security-classes-09}
P.~Urien, ``Security classes for iot devices.''
  https://datatracker.ietf.org/doc/draft-urien-lwig-security-classes/09/, Dec.
  2022.
\newblock Work in Progress. Accessed: 2023-3-24.

\bibitem{Malik2019}
M.~Malik, M.~Dutta, and J.~Granjal, ``A survey of key bootstrapping protocols
  based on public key cryptography in the internet of things,'' {\em IEEE
  Access}, vol.~7, pp.~27443--27464, 2019.

\bibitem{delfs2015introduction}
H.~Delfs and H.~Knebl, {\em Introduction to Cryptography: Principles and
  Applications}.
\newblock Springer, 2015.

\bibitem{Mavroeidis2018}
V.~Mavroeidis, K.~Vishi, M.~D., and A.~J{\o}sang, ``The impact of quantum
  computing on present cryptography,'' {\em International Journal of Advanced
  Computer Science and Applications}, vol.~9, no.~3, 2018.

\bibitem{Baraban2010}
M.~Baraban, N.~E. Bonesteel, and S.~H. Simon, ``Resources required for
  topological quantum factoring,'' {\em Physical Review A - Atomic, Molecular,
  and Optical Physics}, vol.~81, 2010.

\bibitem{Roetteler2017}
M.~Roetteler, M.~Naehrig, K.~M. Svore, and K.~Lauter, ``Quantum resource
  estimates for computing elliptic curve discrete logarithms,'' in {\em
  Advances in Cryptology -- ASIACRYPT 2017}, (Cham), pp.~241--270, Springer
  International Publishing, 2017.

\bibitem{Saxena2021}
A.~Saxena, A.~Shukla, and A.~Pathak, ``A hybrid scheme for prime factorization
  and its experimental implementation using ibm quantum processor,'' {\em
  Quantum Information Processing}, vol.~20, p.~112, 2021.

\bibitem{Bocharov2016}
A.~Bocharov, M.~Roetteler, and K.~M. Svore, ``Factoring with qutrits: Shor's
  algorithm on ternary and metaplectic quantum architectures,'' {\em Physical
  Review A}, vol.~96, no.~1, p.~012306, 2017.

\bibitem{Larasati2021}
H.~T. Larasati and H.~Kim, ``Quantum cryptanalysis landscape of shor's
  algorithm for elliptic curve discrete logarithm problem,'' in {\em
  Information Security Applications} (H.~Kim, ed.), (Cham), pp.~91--104,
  Springer International Publishing, 2021.

\bibitem{Grassl2016}
M.~Grassl, B.~Langenberg, M.~Roetteler, and R.~Steinwandt, ``Applying
  grover’s algorithm to aes: quantum resource estimates,'' in {\em
  Post-Quantum Cryptography: 7th International Workshop, PQCrypto 2016,
  Fukuoka, Japan, February 24-26, 2016, Proceedings 7}, pp.~29--43, Springer,
  2016.

\bibitem{10.1007/978-3-662-53008-5_8}
M.~Kaplan, G.~Leurent, A.~Leverrier, and M.~ Naya-Plasencia, ``Breaking
  symmetric cryptosystems using quantum period finding,'' in {\em Advances in
  Cryptology -- CRYPTO 2016}, pp.~207--237, Springer Berlin Heidelberg, 2016.

\bibitem{10.1007/978-3-030-34578-5_20}
X.~Bonnetain, A.~Hosoyamada, M.~Naya-Plasencia, Y.~Sasaki, and
  A.~Schrottenloher, ``Quantum attacks without superposition queries: The
  offline simon's algorithm,'' in {\em Advances in Cryptology -- ASIACRYPT
  2019}, pp.~552--583, Springer International Publishing, 2019.

\bibitem{IBM2021}
IBM, ``Ibm unveils breakthrough 127-qubit quantum processor.''
  https://newsroom.ibm.com/2021-11-16-IBM-Unveils-Breakthrough-127-Qubit-Quantum-Processor,
  Nov 2021.
\newblock Accessed: 2023-3-24.

\bibitem{IBM2020}
IBM, ``Ibm’s roadmap for scaling quantum technology.''
  https://www.ibm.com/quantum/roadmap, 2020.
\newblock Accessed: 2023-3-24.

\bibitem{Mosca2018}
M.~Mosca, ``Cybersecurity in an era with quantum computers: Will we be
  ready?,'' {\em IEEE Security \& Privacy}, vol.~16, pp.~38--41, 2018.

\bibitem{Bernstein2017}
D.~J. Bernstein and T.~Lange, ``Post-quantum cryptography,'' {\em Nature},
  vol.~549, pp.~188--194, Oct 2017.

\bibitem{nccoe_2021}
NCCoE, ``Migration to post-quantum cryptography.''
  https://www.nccoe.nist.gov/crypto-agility-considerations-migrating-post-quantum-cryptographic-algorithms,
  2021.
\newblock Accessed: 2023-3-24.

\bibitem{nist-sp800-208}
D.~A. Cooper, D.~C. Apon, Q.~H. Dang, M.~S. Davidson, M.~J. Dworkin, C.~A.
  Miller, {\em et~al.}, ``Recommendation for stateful hash-based signature
  schemes,'' {\em NIST Special Publication}, vol.~800, p.~208, Oct 2020.

\bibitem{9679412}
T.~Prantl, D.~Prantl, L.~Beierlieb, L.~Iffländer, A.~Dmitrienko, S.~Kounev,
  and C.~Krupitzer, ``Performance evaluation for a post-quantum public-key
  cryptosystem,'' in {\em 2021 IEEE International Performance, Computing, and
  Communications Conference (IPCCC)}, pp.~1--7, Oct 2021.

\bibitem{9744589}
J.~Señor, J.~Portilla, and G.~Mujica, ``Analysis of the ntru post-quantum
  cryptographic scheme in constrained iot edge devices,'' {\em IEEE Internet of
  Things Journal}, vol.~9, pp.~18778--18790, Oct 2022.

\bibitem{10.1145/3507657.3529652}
K.~Hines, M.~Raavi, J.-M. Villeneuve, S.~Wuthier, J.~Moreno-Colin, Y.~Bai, and
  S.-Y. Chang, ``Post-quantum cipher power analysis in lightweight devices,''
  in {\em Proceedings of the 15th ACM Conference on Security and Privacy in
  Wireless and Mobile Networks}, WiSec '22, pp.~282--284, Association for
  Computing Machinery, 2022.

\bibitem{8972389}
K.~Shafique, B.~A. Khawaja, F.~Sabir, S.~Qazi, and M.~Mustaqim, ``Internet of
  things (iot) for next-generation smart systems: A review of current
  challenges, future trends and prospects for emerging 5g-iot scenarios,'' {\em
  IEEE Access}, vol.~8, pp.~23022--23040, Jan 2020.

\bibitem{9191980}
K.~Mayes, ``Performance evaluation and optimisation for kyber on the multos iot
  trust-anchor,'' in {\em 2020 IEEE International Conference on Smart Internet
  of Things (SmartIoT)}, pp.~1--8, Aug 2020.

\bibitem{9787987}
C.~Sajimon, K.~Jain, and P.~Krishnan, ``Analysis of post-quantum cryptography
  for internet of things,'' in {\em 2022 6th International Conference on
  Intelligent Computing and Control Systems (ICICCS)}, pp.~387--394, May 2022.

\bibitem{10.1145/3498361.3538766}
C.-C. Chung, C.-C. Pai, F.-S. Ching, C.~Wang, and L.-J. Chen, ``When
  post-quantum cryptography meets the internet of things: An empirical study,''
  in {\em Proceedings of the 20th Annual International Conference on Mobile
  Systems, Applications and Services}, MobiSys '22, (New York, NY, USA),
  p.~525–526, Association for Computing Machinery, 2022.

\bibitem{10.1145/3320269.3384725}
K.~B\"{u}rstinghaus-Steinbach, C.~Krau\ss{}, R.~Niederhagen, and M.~Schneider,
  ``Post-quantum tls on embedded systems: Integrating and evaluating kyber and
  sphincs+ with mbed tls,'' in {\em Proceedings of the 15th ACM Asia Conference
  on Computer and Communications Security}, ASIA CCS '20, (New York, NY, USA),
  pp.~841--852, Association for Computing Machinery, 2020.

\bibitem{cryptoeprint:2022/1712}
R.~Gonzalez and T.~Wiggers, ``Kemtls vs. post-quantum tls: Performance on
  embedded systems.'' Cryptology ePrint Archive, Paper 2022/1712, 2022.

\bibitem{10.1145/3450268.3453528}
M.~Sch\"{o}ffel, F.~Lauer, C.~C. Rheinl\"{a}nder, and N.~Wehn, ``On the energy
  costs of post-quantum kems in tls-based low-power secure iot,'' in {\em
  Proceedings of the International Conference on Internet-of-Things Design and
  Implementation}, IoTDI '21, (New York, NY, USA), pp.~158--168, Association
  for Computing Machinery, 2021.

\bibitem{multos_2022}
MULTOS, ``The multos trust anchor development board.''
  https://multos.com/support/multos-trust-anchor/developer-boards/.
\newblock Accessed: 2022-3-24.

\bibitem{10.1145/3372297.3423350}
P.~Schwabe, D.~Stebila, and T.~Wiggers, ``Post-quantum tls without handshake
  signatures,'' in {\em Proceedings of the 2020 ACM SIGSAC Conference on
  Computer and Communications Security}, CCS '20, (New York, NY, USA),
  p.~1461–1480, Association for Computing Machinery, 2020.

\bibitem{9217806}
J.~Winkler, A.~Höller, and C.~Steger, ``Optimizing picnic for limited memory
  resources,'' in {\em 2020 23rd Euromicro Conference on Digital System Design
  (DSD)}, pp.~200--204, Aug 2020.

\bibitem{cryptoeprint:2022/323}
J.~W. Bos, J.~Renes, and A.~Sprenkels, ``Dilithium for memory constrained
  devices.'' Cryptology ePrint Archive, Paper 2022/323, 2022.

\bibitem{9682594}
J.~Zhang, J.~Huang, Z.~Liu, and S.~S. Roy, ``Time-memory trade-offs for saber+
  on memory-constrained risc-v platform,'' {\em IEEE Transactions on
  Computers}, vol.~71, pp.~2996--3007, Nov 2022.

\bibitem{fft-algorithm}
J.~W. Cooley and J.~W. Tukey, ``An algorithm for the machine calculation of
  complex fourier series,'' {\em Mathematics of computation}, vol.~19, no.~90,
  pp.~297--301, 1965.

\bibitem{cryptoeprint:2020/1216}
E.~Alkim, D.~Y.-L. Cheng, C.-M.~M. Chung, H.~Evkan, L.~W.-L. Huang, V.~Hwang,
  C.-L.~T. Li, R.~Niederhagen, C.-J. Shih, J.~Wälde, and B.-Y. Yang,
  ``Polynomial multiplication in ntru prime: Comparison of optimization
  strategies on cortex-m4.'' Cryptology ePrint Archive, Paper 2020/1216, 2020.

\bibitem{good_1951}
I.~J. Good, ``Random motion on a finite abelian group,'' {\em Mathematical
  Proceedings of the Cambridge Philosophical Society}, vol.~47, no.~4,
  p.~756–762, 1951.

\bibitem{cryptoeprint:2020/898}
J.~W. Bos, A.~Hülsing, J.~Renes, and C.~van Vredendaal, ``Rapidly verifiable
  xmss signatures.'' Cryptology ePrint Archive, Paper 2020/898, 2020.

\bibitem{9762260}
Y.~Kim, J.~Song, and S.~C. Seo, ``Accelerating falcon on armv8,'' {\em IEEE
  Access}, vol.~10, pp.~44446--44460, 2022.

\bibitem{cryptoeprint:2021/1117}
H.~Seo, H.~Kwon, S.~Eum, K.~Jang, H.~Kim, H.~Kim, M.~Sim, G.~Song, and W.-K.
  Lee, ``All the polynomial multiplication you need on risc-v.'' Cryptology
  ePrint Archive, Paper 2021/1117, 2021.

\bibitem{cryptoeprint:2020/1303}
J.~W. Bos, J.~Renes, and C.~van Vredendaal, ``Post-quantum cryptography with
  contemporary co-processors: Beyond kronecker, schönhage-strassen \&
  nussbaumer.'' Cryptology ePrint Archive, Paper 2020/1303, 2020.

\bibitem{10.1145/3564625.3564629}
J.~Zheng, F.~He, S.~Shen, C.~Xue, and Y.~Zhao, ``Parallel small polynomial
  multiplication for dilithium: A faster design and implementation,'' in {\em
  Proceedings of the 38th Annual Computer Security Applications Conference},
  ACSAC '22, (New York, NY, USA), pp.~304--317, Association for Computing
  Machinery, 2022.

\bibitem{s22197496}
A.~Magyari and Y.~Chen, ``Review of state-of-the-art fpga applications in iot
  networks,'' {\em Sensors}, vol.~22, no.~19, 2022.

\bibitem{9001747}
M.~Elnawawy, A.~Farhan, A.~A. Nabulsi, A.~Al-Ali, and A.~Sagahyroon, ``Role of
  fpga in internet of things applications,'' in {\em 2019 IEEE International
  Symposium on Signal Processing and Information Technology (ISSPIT)},
  pp.~1--6, 2019.

\bibitem{10.1145/2899007.2899011}
J.~Buchmann, F.~G\"{o}pfert, T.~G\"{u}neysu, T.~Oder, and T.~P\"{o}ppelmann,
  ``High-performance and lightweight lattice-based public-key encryption,'' in
  {\em Proceedings of the 2nd ACM International Workshop on IoT Privacy, Trust,
  and Security}, IoTPTS '16, (New York, NY, USA), p.~2–9, Association for
  Computing Machinery, 2016.

\bibitem{9211858}
S.~Ebrahimi and S.~Bayat-Sarmadi, ``Lightweight and dpa-resistant post-quantum
  cryptoprocessor based on binary ring-lwe,'' in {\em 2020 20th International
  Symposium on Computer Architecture and Digital Systems (CADS)}, pp.~1--6, Aug
  2020.

\bibitem{9717291}
S.~Hadayeghparast, S.~Bayat-Sarmadi, and S.~Ebrahimi, ``High-speed post-quantum
  cryptoprocessor based on risc-v architecture for iot,'' {\em IEEE Internet of
  Things Journal}, vol.~9, pp.~15839--15846, Sep. 2022.

\bibitem{9737700}
B.~J. Lucas, A.~Alwan, M.~Murzello, Y.~Tu, P.~He, A.~J. Schwartz, D.~Guevara,
  U.~Guin, K.~Juretus, and J.~Xie, ``Lightweight hardware implementation of
  binary ring-lwe pqc accelerator,'' {\em IEEE Computer Architecture Letters},
  vol.~21, no.~1, pp.~17--20, 2022.

\bibitem{10.1145/3569457}
P.~He, T.~Bao, J.~Xie, and M.~Amin, ``Fpga implementation of compact hardware
  accelerators for ring-binary-lwe based post-quantum cryptography,'' {\em ACM
  Trans. Reconfigurable Technol. Syst.}, oct 2022.
\newblock Just Accepted.

\bibitem{8342207}
A.~Aysu, M.~Orshansky, and M.~Tiwari, ``Binary ring-lwe hardware with power
  side-channel countermeasures,'' in {\em 2018 Design, Automation \& Test in
  Europe Conference \& Exhibition (DATE)}, pp.~1253--1258, 2018.

\bibitem{9911531}
B.~Kim, J.~Park, S.~Moon, K.~Kang, and J.-Y. Sim, ``Configurable
  energy-efficient lattice-based post-quantum cryptography processor for iot
  devices,'' in {\em ESSCIRC 2022- IEEE 48th European Solid State Circuits
  Conference (ESSCIRC)}, pp.~525--528, Sep. 2022.

\bibitem{cryptoeprint:2022/217}
L.~Beckwith, D.~T. Nguyen, and K.~Gaj, ``High-performance hardware
  implementation of lattice-based digital signatures.'' Cryptology ePrint
  Archive, Paper 2022/217, 2022.

\bibitem{9460703}
Q.~Berthet, A.~Upegui, L.~Gantel, A.~Duc, and G.~Traverso, ``An area-efficient
  sphincs+ post-quantum signature coprocessor,'' in {\em 2021 IEEE
  International Parallel and Distributed Processing Symposium Workshops
  (IPDPSW)}, pp.~180--187, June 2021.

\bibitem{9180839}
D.~E.~S. Kundi, S.~Bian, A.~Khalid, C.~Wang, M.~O'Neill, and W.~Liu, ``Axmm:
  Area and power efficient approximate modular multiplier for r-lwe
  cryptosystem,'' in {\em 2020 IEEE International Symposium on Circuits and
  Systems (ISCAS)}, pp.~1--5, Oct 2020.

\bibitem{8877876}
J.~Hu, M.~Baldi, P.~Santini, N.~Zeng, S.~Ling, and H.~Wang, ``Lightweight key
  encapsulation using ldpc codes on fpgas,'' {\em IEEE Transactions on
  Computers}, vol.~69, pp.~327--341, March 2020.

\bibitem{9089230}
J.~Hu, Y.~Liu, R.~C.~C. Cheung, S.~Bhasin, S.~Ling, and H.~Wang, ``Compact
  code-based signature for reconfigurable devices with side channel
  resilience,'' {\em IEEE Transactions on Circuits and Systems I: Regular
  Papers}, vol.~67, pp.~2305--2316, July 2020.

\bibitem{9712707}
M.~Xin, C.~Xu, K.~Huang, H.~Yu, H.~Yao, X.~Jiang, and D.~Liu, ``Implementation
  of number theoretic transform unit for polynomial multiplication of
  lattice-based cryptography,'' in {\em 2022 2nd International Conference on
  Consumer Electronics and Computer Engineering (ICCECE)}, pp.~323--327, Jan
  2022.

\bibitem{9559627}
Y.~Cao, Y.~Wu, W.~Wang, X.~Lu, S.~Chen, J.~Ye, and C.-H. Chang, ``An efficient
  full hardware implementation of extended merkle signature scheme,'' {\em IEEE
  Transactions on Circuits and Systems I: Regular Papers}, vol.~69,
  pp.~682--693, Feb 2022.

\bibitem{9869764}
Y.~Cao, Y.~Wu, L.~Qin, S.~Chen, and C.-H. Chang, ``Area, time and energy
  efficient multicore hardware accelerators for extended merkle signature
  scheme,'' {\em IEEE Transactions on Circuits and Systems I: Regular Papers},
  vol.~69, pp.~4908--4918, Dec 2022.

\bibitem{10.1007/978-3-540-85053-3_6}
R.~Szerwinski and T.~G{\"u}neysu, ``Exploiting the power of gpus for asymmetric
  cryptography,'' in {\em Cryptographic Hardware and Embedded Systems -- CHES
  2008}, pp.~79--99, Springer Berlin Heidelberg, 2008.

\bibitem{su14148708}
M.~Shuaib, S.~Badotra, M.~I. Khalid, A.~D. Algarni, S.~S. Ullah, S.~Bourouis,
  J.~Iqbal, S.~Bharany, and L.~Gundaboina, ``A novel optimization for gpu
  mining using overclocking and undervolting,'' {\em Sustainability}, vol.~14,
  no.~14, 2022.

\bibitem{9860310}
W.-K. Lee and S.~O. Hwang, ``High throughput implementation of post-quantum key
  encapsulation and decapsulation on gpu for internet of things applications,''
  in {\em 2022 IEEE World Congress on Services (SERVICES)}, pp.~13--13, July
  2022.

\bibitem{9667701}
K.~Lee, M.~Gowanlock, and B.~Cambou, ``Saber-gpu: A response-based cryptography
  algorithm for saber on the gpu,'' in {\em 2021 IEEE 26th Pacific Rim
  International Symposium on Dependable Computing (PRDC)}, pp.~123--132, Dec
  2021.

\bibitem{9715141}
W.-K. Lee, H.~Seo, Z.~Zhang, and S.~O. Hwang, ``Tensorcrypto: High throughput
  acceleration of lattice-based cryptography using tensor core on gpu,'' {\em
  IEEE Access}, vol.~10, pp.~20616--20632, 2022.

\bibitem{cryptoeprint:2021/1389}
W.-K. Lee, H.~Seo, S.~O. Hwang, A.~Karmakar, J.~M.~B. Mera, and R.~Achar,
  ``Dpcrypto: Acceleration of post-quantum cryptographic algorithms using
  dot-product instruction on gpus.'' Cryptology ePrint Archive, Paper
  2021/1389, 2021.

\bibitem{9511134}
W.-K. Lee and S.~O. Hwang, ``High throughput implementation of post-quantum key
  encapsulation and decapsulation on gpu for internet of things applications,''
  {\em IEEE Transactions on Services Computing}, vol.~15, pp.~3275--3288, Nov
  2022.

\end{thebibliography}

%

\begin{IEEEbiography}[{\includegraphics[width=1in,height=1.25in,clip,keepaspectratio]{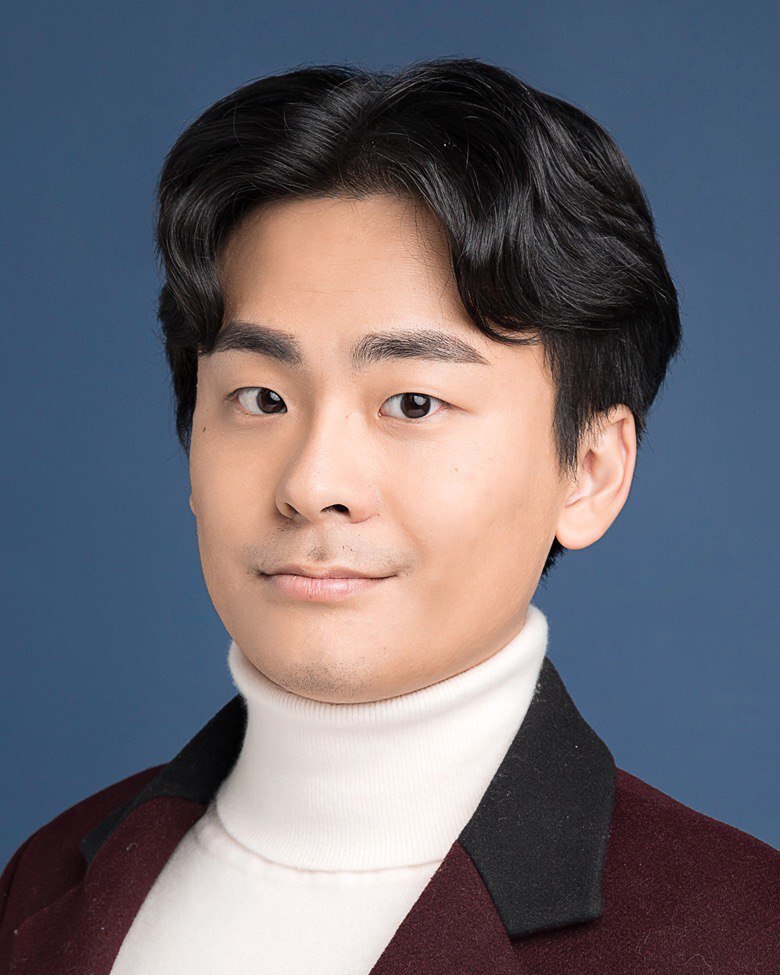}}]{Tao Liu}
Received his M. IT degree from Queensland University of Technology in 2022. He is currently pursuing a PhD Degree in computer science with QUT. His research interest include the Internet-of-Things, cybersecurity and data encryption. 
\end{IEEEbiography}


\begin{IEEEbiography}[{\includegraphics[width=1in,height=1.25in,clip,keepaspectratio]{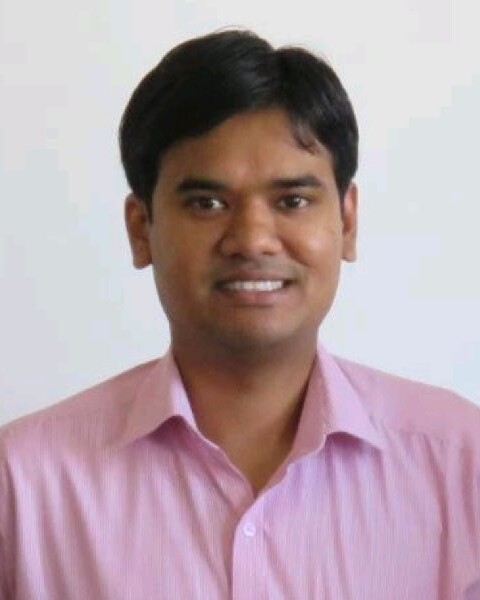}}]{Gowri Sanker Ramachandran}
received the M.Sc. degree from Malardalen University, Sweden, and the Ph.D. degree from KU Leuven, Belgium. He is a Research Fellow of Distributed Systems, Blockchain, and the Internet of Things with the Queensland University of Technology, Australia. He was a Postdoctoral Researcher and a Senior Research Associate with the University of Southern California, USA, from 2017 to 2020. He has published over 50 peer-reviewed publications. His research interests include blockchain, IoT, and distributed computing. He serves on the organizing committees of top international conferences, including ICBC, IPSN, IoTDI, and SenSys. He was the General Vice-Chair of ACM BlockSys-2022, a co-located workshop with SenSys-2022.
\end{IEEEbiography}

\newpage

\begin{IEEEbiography}[{\includegraphics[width=1in,height=1.25in,clip,keepaspectratio]{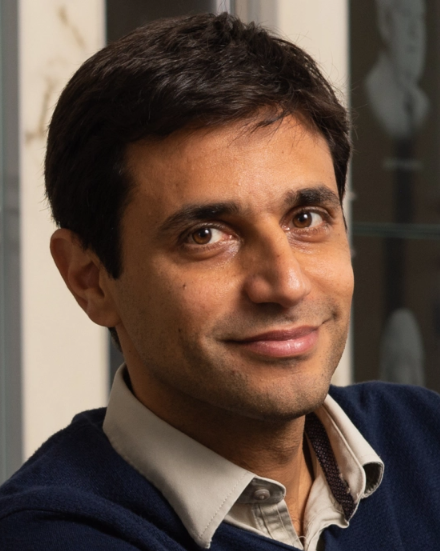}}]{Raja Jurdak}
received the M.S. and Ph.D. degrees from the University of California at Irvine. He is a Professor of Distributed Systems and the
Chair of Applied Data Sciences with the Queensland University of Technology, and the Director of the Trusted Networks Lab. He previously established and led the Distributed Sensing Systems Group, Data61, CSIRO. He also spent time as a Visiting Academic with MIT and Oxford University in 2011 and 2017. He is a Conjoint Professor with the University of New South Wales and a Visiting Researcher with Data61, CSIRO. He has published over 230 peer-reviewed publications, including two authored books most recently on blockchain in cyber–physical systems in 2020. His publications have attracted over 11,000 citations, with an H-index of 47. His research interests include blockchain, IoT, trust, mobility, and energy efficiency in networks. He serves on the editorial boards of Ad Hoc Networks and Scientific Reports (Nature), and on the organizing and technical program committees of top international conferences, including Percom, ICBC, IPSN, WoWMoM, and ICDCS. He was the TPC Co-Chair of ICBC in 2021.
\end{IEEEbiography}




\end{document}